\documentclass[manuscript]{aastex}
\usepackage{emulateapj5}
\usepackage{apjfonts}
\usepackage{times}
\usepackage{natbib}

\makeatletter

\newenvironment{inlinefigure}{%
\def\@captype{figure}%
\noindent\begin{minipage}{0.999\linewidth}\begin{center}}
{\end{center}\end{minipage}\smallskip}
\makeatother

\usepackage{graphicx}
\shorttitle{Removing foreground using non-Gaussianity}

\begin{document}

\title{A Foreground Cleaned CMB Map from Non-Gaussianity Measurement} 

\author{Rajib Saha\altaffilmark{1}}

 \altaffiltext{1}{Physics Department, Indian Institute of Science Education and Research Bhopal,
 Bhopal, M.P, 462023, India.} 

\begin{abstract}
In this paper we present a new method to estimate a foreground cleaned 
Cosmic Microwave  Background (CMB) map at a resolution of $1^\circ$ by 
minimizing the non-Gaussian properties of the cleaned map  which arise  
dominantly due to diffuse foreground emission components from the Milky Way. 
We employ simple kurtosis statistic as the measure of non-Gaussian 
properties and perform  a linear combination of $5$ frequency maps provided 
by Wilkinson Microwave Anisotropy Probe (WMAP) in its $7$ year data release 
in such a way that the cleaned map has a minimum kurtosis which leads  
to a non-Gaussianity minimized, foreground cleaned CMB map. We validate the 
method  by performing Monte-Carlo simulations. To minimize any residual 
foreground contamination from the cleaned map we flag out region near the galactic 
plane  based upon results from simulations. Outside the masked region our new estimate 
of CMB map matches well with the WMAP's ILC map. A simple pseudo-$C_l$  based CMB 
TT  power spectrum derived from the non-gaussianity minimized map  reproduces the earlier 
results of WMAP's power spectrum. {\it An important advantage of the method is that it does 
not introduce  any negative bias in angular power spectrum in low multipole regime, unlike 
usual ILC method.} Comparing our results with the previously published results we argue 
that CMB results are robust with respect to specific
foreground removal algorithms employed.
\end{abstract}


\section{Introduction}
The Cosmic Microwave Background (CMB) has conceivably become the finest 
tool so far to probe the physics of the early universe.  The problem of  
isolating a clean CMB signal from contaminations originating from Milky Way 
is of primary importance to a cosmologist. 
According to the slowly rolling single scalar field inflationary scenario the primordial perturbations
follow Gaussianity to a very good approximation, any non Gaussian properties 
predicted to be a small effect~(\cite{Allen87,Falk92,Gangui94,Acquaviva03,Maldacena03}). Since CMB  anisotropies are directly related to these
perturbations via the spacetime metric one expects CMB  to follow Gaussian distribution. 
It was shown by~\cite{Munshi95} that non-Gaussian effect introduced in CMB  
after decoupling is also small. It has been shown by~\cite{Komatsu03} that
CMB data observed by WMAP satellite~(\cite{Bennett97}) is consistent with primordial 
Gaussian fluctuation. Contrary to these, diffuse foreground components,
originating from our own galaxy exhibit highly non-Gaussian properties due to 
non-linearities involved in the physical processes during their origin.  Based upon  simple assumption 
of pure Gaussian properties of CMB and non-Gaussian nature 
of diffuse foregrounds, in this paper, we propose a new foreground removal method from CMB maps. A detailed 
account on several other foreground removal  methods may be found in existing 
literatures~(\cite{Brandt94,Tegmark96,Bouchet99,Bennett03,Tegmark03,Saha06,Hansen06,Eriksen06,Hinshaw07,Saha08,Eriksen08a,
Eriksen08b,Leach08,Delabrouille09,Samal10,Pietrobon10}). For discussions about
searching for non-Gaussianity in foreground minimized CMB maps  
we refer to~\cite{Rath09,Barreiro2000,Banday2000,Raeth10,Hou10,Bernui10}. 
    
\section{Methodology}
\label{method}
\subsection{Estimator for non-Gausianity}
 A measure of non-Gaussian properties  inherent in a collection of $N$ random samples is given 
by so called kurtosis statistic, ($\mathcal K$).  
For a set, $S$, of $N$ random samples,  $S = \{x_i \vert i = 1, 2, 3, ..., N\}$, the 
$\mathcal K$ statistic is defined as, 
\begin{eqnarray}
\mathcal K = \frac{1}{N}\sum_{i=1}^{i=N}\frac{\left (x_i-x_0 \right)^4}{\sigma^4} - 3\, , 
\label{K}
\end{eqnarray} 
where $x_0$ denotes sample mean and $\sigma$ is the standard deviation of samples.   
$\mathcal K = 0$ for samples drawn from Gaussian distribution whereas as discussed in Section~\ref{Kurt} 
 foreground distributions possess large positive kurtosis.

\subsection{Foreground Cleaned Map}
Let ${\bf X}_f$ denotes foreground contaminated CMB map in thermodynamic temperature unit at $1^\circ$ resolution 
at a frequency index $f$  and  $n_b$ represents  total number of available 
frequency bands. ${\bf X}_f$ is a column vector of $n$ numbers where $n$ is the number
of surviving pixels  at a frequency band after masking the known point source positions. 
We form a foreground  cleaned map  following,  
\begin{eqnarray}
{\bf X}^c = \sum_{f=1}^{n_b} w_f {\bf X}_f\, ,
\label{lc}
\end{eqnarray}
where $w_f$ denotes the weight  factor for linear combination for frequency index $f$. We choose these factors 
such that $\mathcal K$ value for ${\bf X}^c$ is minimized with an imposed constraint on them that 
$\sum_{f=1}^{n_b} w_f = 1$. This condition preserves the actual values of CMB temperature at each pixel in the 
foreground cleaned map. 

Using Eqns.~\ref{K} and~\ref{lc} one can show that, $\mathcal K^c({\bf W})$, of the linearly combined  map follows, 
\begin{eqnarray}
\mathcal K^c({\bf W}) = \left [ \frac{n}{\left ({\bf W T W}^T \right)^2} \sum_{i=1}^{n}\left( {\bf W T}_j{\bf W}^T\right)^2 \right ] - 3\, ,
\end{eqnarray} 
where $\bf W$ denotes a $1 \times n_b$ vector whose $f^{th}$ entry is given by $w_f$ and ${\bf T} = \sum_{j=1}^{n} {\bf T}_j$.
Here ${\bf T}_j$ is a $n_b \times n_b$ symmetric matrix for pixel $j$,  
${\bf T}_{j(ff')} = \Delta \bar T_{fj} \Delta \bar T_{f'j}$ where $\Delta \bar T_{fj}$  
denotes the temperature at pixel $j$ of frequency index $f$ after the mean temperature corresponding to this frequency 
has been subtracted from actual pixel temperature. 

 In principle, the solution for $\bf W$ for minimum $\mathcal K^c({\bf W})$ satisfying our conditions 
can be obtained by employing a Lagrange's undetermined multiplier approach. However, because of 
nontrivial nature of dependency 
of $\mathcal K^c$ on $\bf W$ we find that such an approach is not feasible for our problem. Instead,  we  
find the minimum of $\mathcal K^c({\bf W})$ 
by invoking a non-linear search algorithm due to Powell.

\section{Results} 
\label{Result}
\subsection{Kurtosis}
\label{Kurt}

\begin{inlinefigure}
\centerline{\includegraphics[scale=0.8]{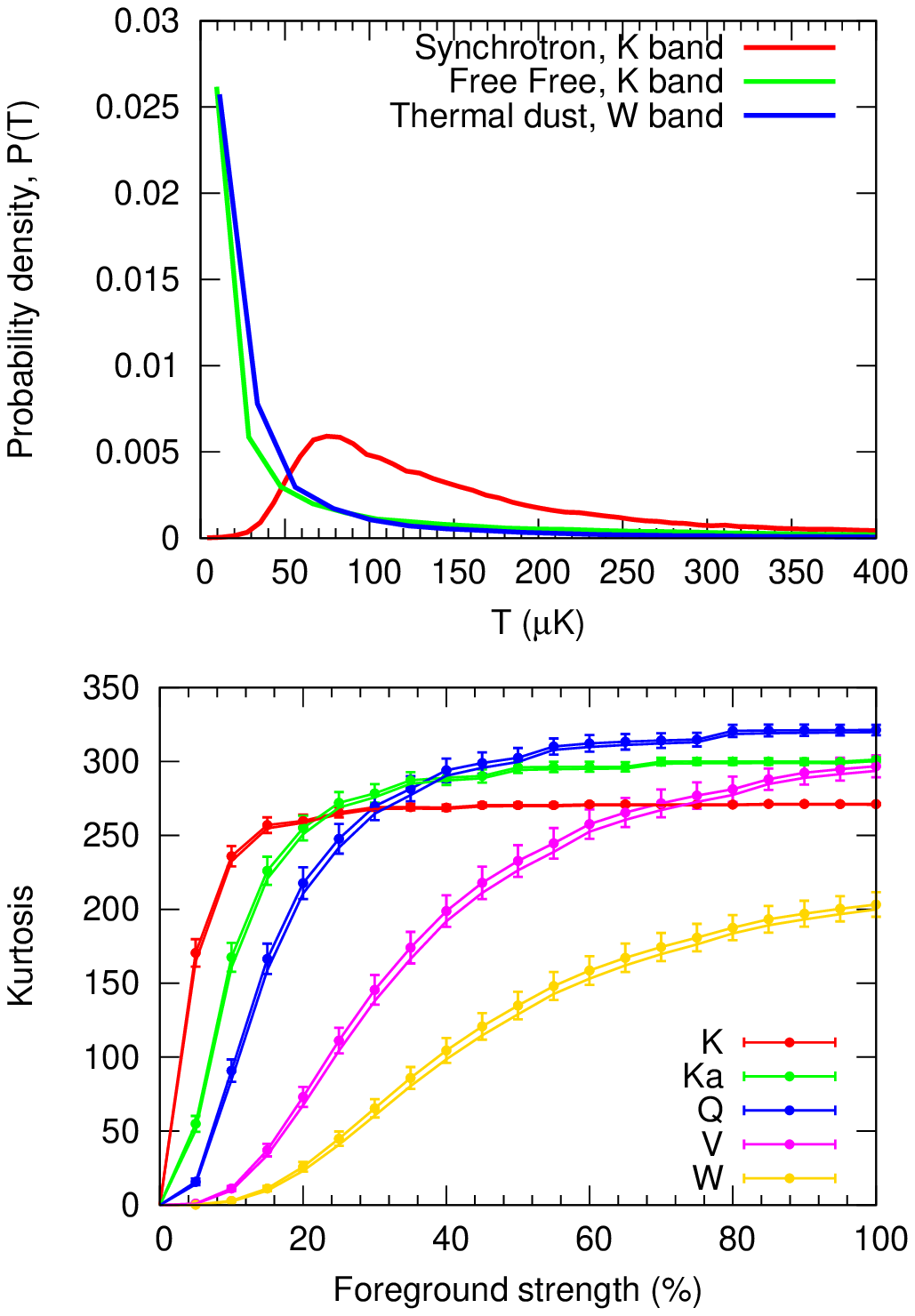}}
\caption{{\it Top Panel:} Non-Gaussian distribution of pixel temperature of MEM foreground
maps. The frequency bands for this plots  are chosen such that respective
foreground component is most dominant over other frequencies inside the WMAP window.
{\it Bottom Panel:} The plot of variation of kurtosis, $\mathcal K$,
as a function of foreground strength at WMAP frequencies. The solid 
lines show mean curves obtained from the $150$ random simulations 
of CMB and foreground with foreground strength as indicated by the 
ordinate of each points. The dashed lines show the variation of 
kurtosis for a randomly  chosen realization. The error bars indicate 
cosmic variance.   Mean kurtosis  estimated from  pure CMB maps  is given by, $\left < \mathcal K_{CMB} \right> = -0.023 \pm 0.060.$\label{Kvals}}
\end{inlinefigure}

The distributions of  pixel temperatures due to  diffuse galactic 
foregrounds are shown in top panel of Fig.~\ref{Kvals}. Each of these foreground
distributions is strongly asymmetric and exhibits a long tail towards 
positive temperature direction indicating 
a variation slower than $\sim e^{-T^2}$, which is the case for a Gaussian
distribution.  The peak for the synchrotron distribution shows that at $23$GHz most likely
contamination  due to synchrotron occurs at a temperature $\sim 75\mu K$,
although contaminations at high ($> 400 \mu K$) and low (e.g., as low as $ < 25\mu K$)
pixel temperature are also likely. 

How does $\mathcal K$  vary with  foreground contamination in  CMB 
maps at different WMAP frequencies? To answer this question we generate $150$ random full sky CMB   
maps using WMAP's LCDM power spectrum. With each of these random realization of CMB map
we add WMAP's MEM foreground templates at various strengths. The resulting behavior 
of the $\mathcal K$ with foreground strength is shown in Fig.~\ref{Kvals} for all WMAP 
frequency bands. For each band  $\mathcal K$ increases as the  amount of foreground 
contamination increases. For K band the variation takes a shape of a plateau, characterized 
by a slow increase of $\mathcal K$,  as the foreground 
level reaches $\sim 20\%$ of its full level. For all bands $\mathcal K$ becomes more than $150$  when all foregrounds 
are operative at their $100\%$ level. It is interesting to note that the Q band (not the K band 
which has the highest level of synchrotron and free free contamination)  shows the 
largest  $\mathcal K$ value among all the $5$ bands  at the maximum foreground strength. 

\subsection{Validation of the Method}
\label{validity}
We validate the methodology  by performing Monte-Carlo simulations.
We use MEM foreground maps for synchrotron, free free and thermal dust
available from LAMBDA website. These maps are provided in a common resolution 
of $1^\circ$ and at pixel resolution parameter $nside = 256$ in antenna 
millikelvin temperature unit. We upgrade the pixel resolution of each map to $nside = 512$ and 
convert them to thermodynamic microkelvin unit. We add the composite foreground 
map of each WMAP frequency with a random  realization of  CMB compatible to LCDM model 
to make foreground contaminated CMB maps at each of the WMAP frequencies. Finally, we mask
the position of the known point sources using the WMAP's 7 year point source mask. 
   
 We develop a C code (hereafter {\it GaussMap}) to implement constrained Powell's conjugate gradient method.
We perform $200$ Monte-Carlo simulations of foreground removal method. From these 
simulations  we find that the foreground removal is effective over almost all parts
of the sky. However, we find visible signature of some residual foreground emission 
in the inner plane of the galaxy. To find out the sky regions where the residual foreground 
could be significant compared to the expected CMB signal we  subtract average 
of input CMB maps from the average of foreground cleaned maps. 
Then we form an initial  mask by assigning zero values to all pixels with absolute temperature
values more than $20\mu$K  of this map and unity at all other pixels. This mask contains 
a set of scattered pixels in the inner galactic plane. To remove these pixels we first 
smooth the initial mask by a Gaussian window of $1^\circ$. We transform this smoothed mask
to a new mask by assigning all pixels of smoothed mask with values greater than or  equal to
$0.9$ to a new value of unity and the all other pixels  to zero. To exclude position of the known  point sources 
from the analysis we make the final mask multiplying this mask by the WMAP's point source
mask. We call the resulting mask as G20 mask. This retains $\sim 87\%$ of the entire sky area.

We estimate $\mathcal K$ for each of the cleaned maps obtained from Monte-Carlo simulations 
of foreground removal procedure~{\footnote{Note that the position of known point sources are excluded 
before computing the $\mathcal K$ values.}}. The mean kurtosis obtained from all the cleaned maps 
is given by $\left < \mathcal K \right> = 0.86 \pm
0.21$.  This corresponds to  $\sim 4\sigma$ detection  of non Gaussianity from the 
cleaned  maps. We interpret the non-vanishing  $\left < \mathcal K \right>$  in terms of residual foreground 
contamination originating from the  galactic plane. After flagging of the pixels determined by the  G20 mask
we obtain, $\left < \mathcal K \right> = -0.02  \pm 0.06$, which is consistent with zero. 
Henceforth, we  use G20 mask as the basic mask to remove pixels contaminated by the residual foreground while 
analyzing cleaned maps from Monte-Carlo simulations as well as WMAP data. The mean weights for $5$ frequency 
bands satisfy, $\left <{\bf W}\right> = (0.049 \pm  0.021, -0.419 \pm   0.063, 
-0.213 \pm   0.027, 1.643 \pm   0.063, -0.059 \pm  0.029)$. 

\subsubsection{Temperature Distribution}

Using the Monte-Carlo simulations  we verify that the pixel temperature of cleaned maps outside the G20 
mask follows a Gaussian distribution. For this we apply G20 mask both to a randomly chosen  CMB realization and its 
foreground cleaned counterpart. 
To verify the Gaussian nature of these distributions we fit the histogram of 
input CMB map (after G20 mask is applied) by a normalized Gaussian probability distribution, $g(T) =
exp(-(T-a)^2/(2s^2))/\sqrt{(2 \pi s^2)}$, where $a$ and $s$  denote respectively mean and 
standard deviation of the distribution. From the fit we find that, 
$s = 69.95 \pm 0.09 \mu K$ and $a = 2.12 \pm 0.11 \mu K$~{\footnote{The 
small monopole is a manifestation of cosmic variance.}}. 

\subsubsection{Power Spectrum}

We apply G20 mask on each of the $200$ foreground cleaned CMB maps and estimate full sky estimate of 
CMB power spectrum using MASTER method~(\cite{Hivon02}). The average of $200$  power spectra obtained from  the foreground 
removed maps matches excellently with the average of the input CMB power spectra. We show  both spectra
in Fig.~\ref{NoBiasCl} along with the cosmic variance.  
This verifies that  no significant foreground contamination exists outside  the G20 mask and the method 
outlined in this paper can be used to estimate CMB  power spectrum  to extract cosmological information.


\begin{inlinefigure}
\centerline{\includegraphics[scale=0.75]{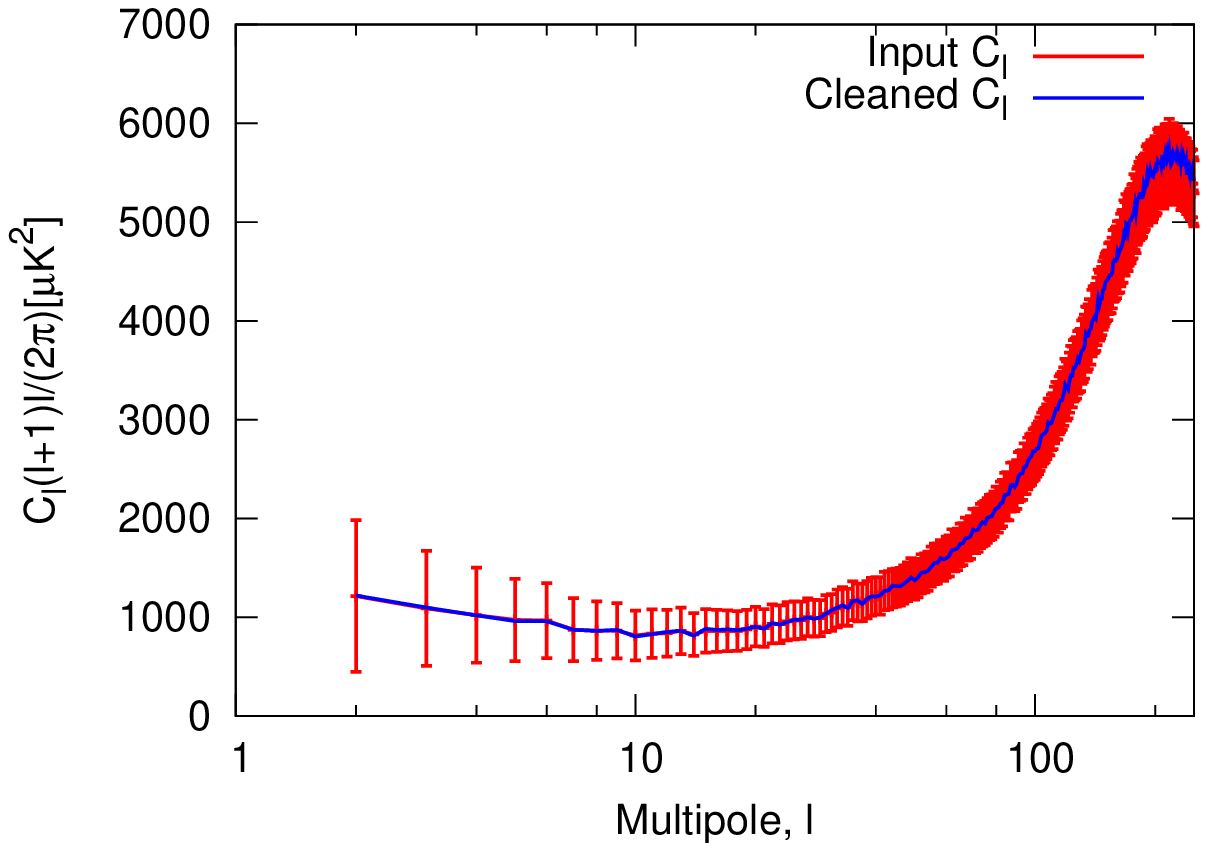}}
\caption{Comparison of average power spectrum obtained from 
foreground cleaned CMB maps outside G20 mask (red) with the 
average input CMB power spectrum, (red line with errorbars). 
The average cleaned  power spectrum is obtained 
by using MASTER algorithm. Both spectra have beam
and pixel window effects removed.\label{NoBiasCl}}
\end{inlinefigure}

 In some of the  earlier publications~(\cite{Saha06,Tarun06,Saha08,Samal10}) the 
authors  reported that the harmonic space based ILC  method gives rise to 
a negative bias in its power spectrum at the low multipoles. The bias appears  due to a mere chance correlation
between  CMB and foregrounds at large angles due to availability of only small number of modes in the large
scales on the sky. In principle, the negative bias in the power spectrum can be present in both pixel based 
and multipole based ILC algorithms which rely upon minimization of net foreground variance from the cleaned map, 
since in these cases the finite correlation between CMB and foregrounds arises due to  certain degree of overlap between 
hot spots and cold spots of CMB and foregrounds and the algorithm creates a negative bias due to nonlinear dependence of weights 
on the  empirical covariance matrix~(\cite{Saha08}). However, the foreground removal method  described in this paper  relies 
upon the Gaussian nature of the final distribution, without explicitly minimizing the variance of the data.
This leads to the advantage that the power spectrum obtained from the cleaned map does not have any negative bias.

\subsection{Application on WMAP data}
Before the  analysis we  mask out the positions of known point sources from each of  WMAP's 
$5$ frequency maps using the point source mask. After masking each map
contains $3054273$ pixels comprising $97\%$ of the full sky area. 

\subsubsection{Non-Gaussianity minimized map}

\begin{inlinefigure}
\centerline{\includegraphics[scale=0.7]{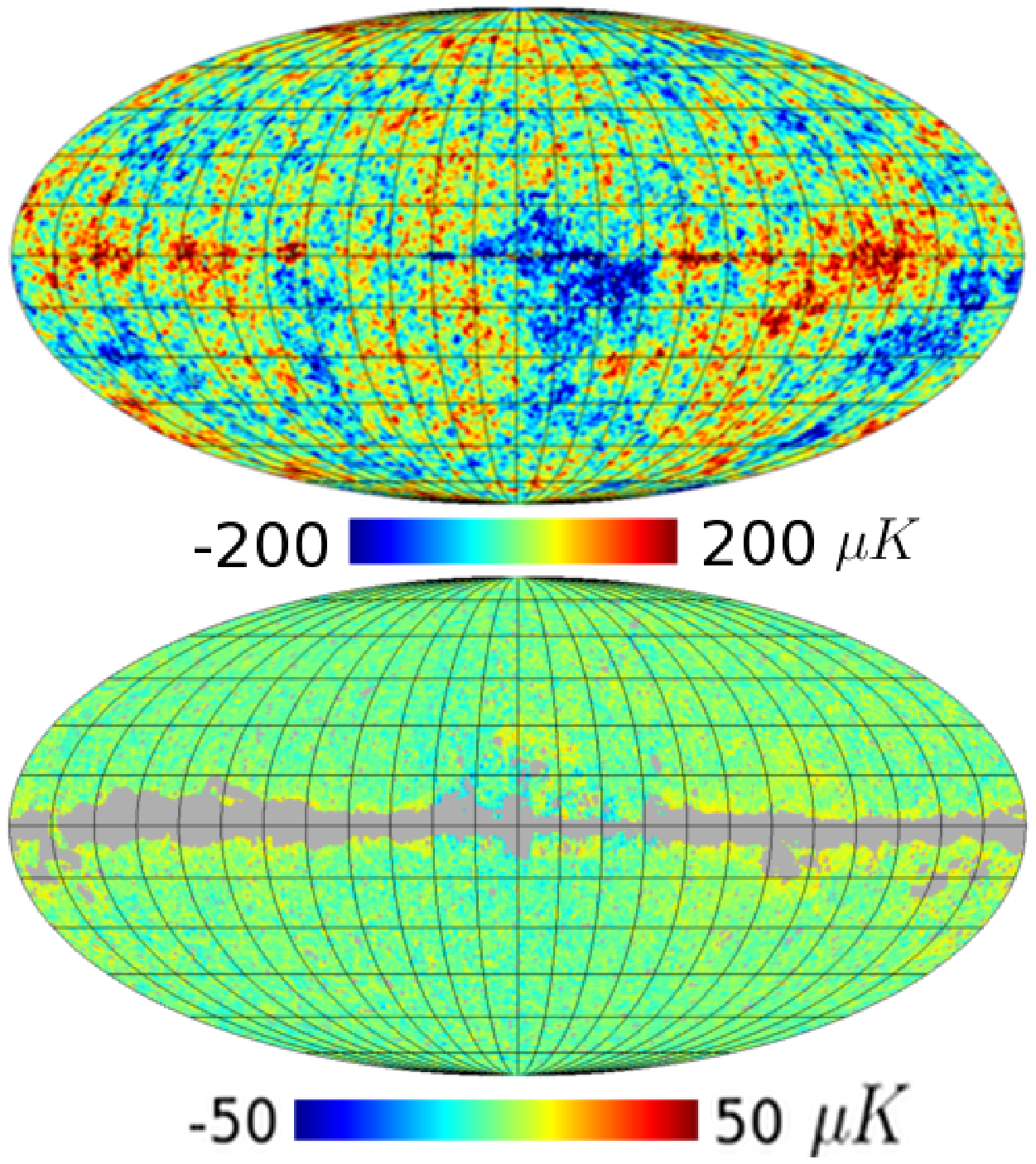}}
\caption{{\it Top panel:} The foreground removed GMAP  obtained by minimizing 
$K^c({\bf W})$. The map has a beam resolution of $1^\circ$ and a pixel 
resolution  corresponding to pixel resolution parameter $nside = 512$.
{\it Bottom panel:} The difference between GMAP 
and the WMAP's $7$ year's ILC map outside G20 mask. \label{Gmap}}
\end{inlinefigure}

We use {\it GaussMap} code to estimate the best fit weight factors corresponding to different frequency bands. The code 
gives best choice of weights as  ${\bf W} = (0.035,  -0.378,  -0.217,  1.632,  -0.073)$ for K to W bands
corresponding to $\mathcal K^c_{min}({\bf W}) = 1.17$. The complete procedure to estimate the weights takes only about 
a couple of minutes on an Intel 2.26 GHz 
processor. As one might expect, V band gets the maximum positive 
weight since it is supposed to be the least foreground contaminated frequency band in the WMAP observation window. The 
kurtosis for cleaned map outside the G20 mask becomes $0.093$. We note that weights obtained from WMAP data have similar values 
to the mean weights  obtained from Monte-Carlo simulations in Section~\ref{validity}.  

Using the weights described above we obtain the foreground cleaned CMB map (top panel of Fig.~\ref{Gmap}, hereafter GMAP). 
Although there are some visible signature of presence of residual foreground contamination in this map on the galactic plane 
(e.g. Cygnus region, Cassiopeia A, Carina nebula) all contaminations are confined only near  the galactic plane. 
We show the difference between GMAP  and WMAP's $7$ year ILC map outside the G20 mask in bottom panel of Fig.~\ref{Gmap}. 
In the unmasked difference map larger pixel amplitudes  are confined near the galactic plane,
making a narrow strip-like structure, with absolute pixel temperature exceeding $50 \mu K$.
The bottom panel of Fig.~\ref{Gmap} shows  that applying G20 mask significantly reduces pixel amplitude of the difference map..

\subsubsection{Temperature Distribution}

\begin{inlinefigure}
\centerline{\includegraphics[scale=0.7]{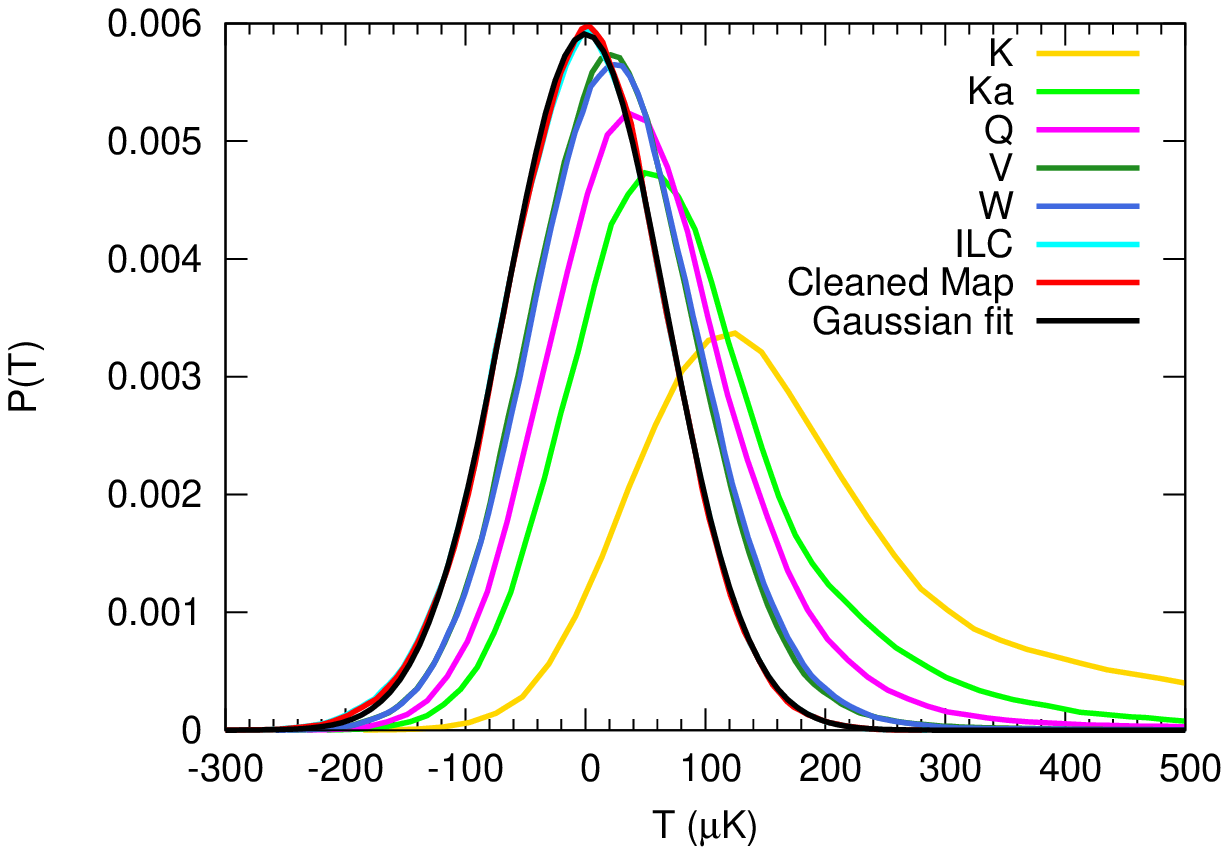}}
\caption{The distribution of the pixel temperatures of GMAP outside 
G20 mask is shown in red. The black curve shows a Gaussian fit to this data.
The cyan curve shows the distribution of ILC map outside G20 mask.
The other curves show the non-Gaussian distributions of pixel temperatures outside G20 mask for each
of the individual WMAP frequency band. Note that since these frequency maps contain CMB their
distributions extend to negative pixel values.\label{res1}}
\end{inlinefigure}

We test  for the distribution of pixel temperature of GMAP 
outside the G20 mask. The average pixel temperature  of this map 
outside the G20 mask  is only $0.85 \mu K$. Hence we fit for a Gaussian 
probability function with mean $0$ and variance $s$ to the  pixel temperature 
distribution of the masked GMAP. From the fit we obtain, $s = 67.49 \pm  0.12 \mu K$.
We show the  distribution outside G20 mask obtained from GMAP and the WMAP's ILC map
in Fig.~\ref{res1}. We also show in this figure pixel temperature distributions of individual 
frequency bands. The long tails for K and Ka bands are due to strong synchrotron contamination. 
We note that all  the distributions corresponding to 5 frequency bands are asymmetric representing their
non-Gaussian nature. Pixel temperature of simulated frequency maps have 
distributions similar to these histograms.

\subsubsection{Power Spectrum}
To estimate power spectrum from GMAP we first apply G20 mask 
and estimate the partial sky CMB power spectrum. We convert the partial sky power spectrum
to the full sky estimate by inverting the mode-mode coupling matrix. Finally we remove 
both beam and pixel effect from this power spectrum. We show the resulting power spectrum 
in Fig.~\ref{cl}. As shown in this figure this power spectrum matches well with the ILC power spectrum estimated 
from the same sky region until $l \sim 100$. Beyond this multipole range 
GMAP power spectrum  has less power than the ILC power spectrum. From the GMAP  we find $C_2 = 246.4 \mu K^2$ and $C_3 = 402.3 \mu K^2$ consistent
with ILC estimates ($C_2 = 248.8 \mu K^2$ and $C_3 = 404.0 \mu K^2$).       


\begin{inlinefigure}
\centerline{\includegraphics[scale=0.75]{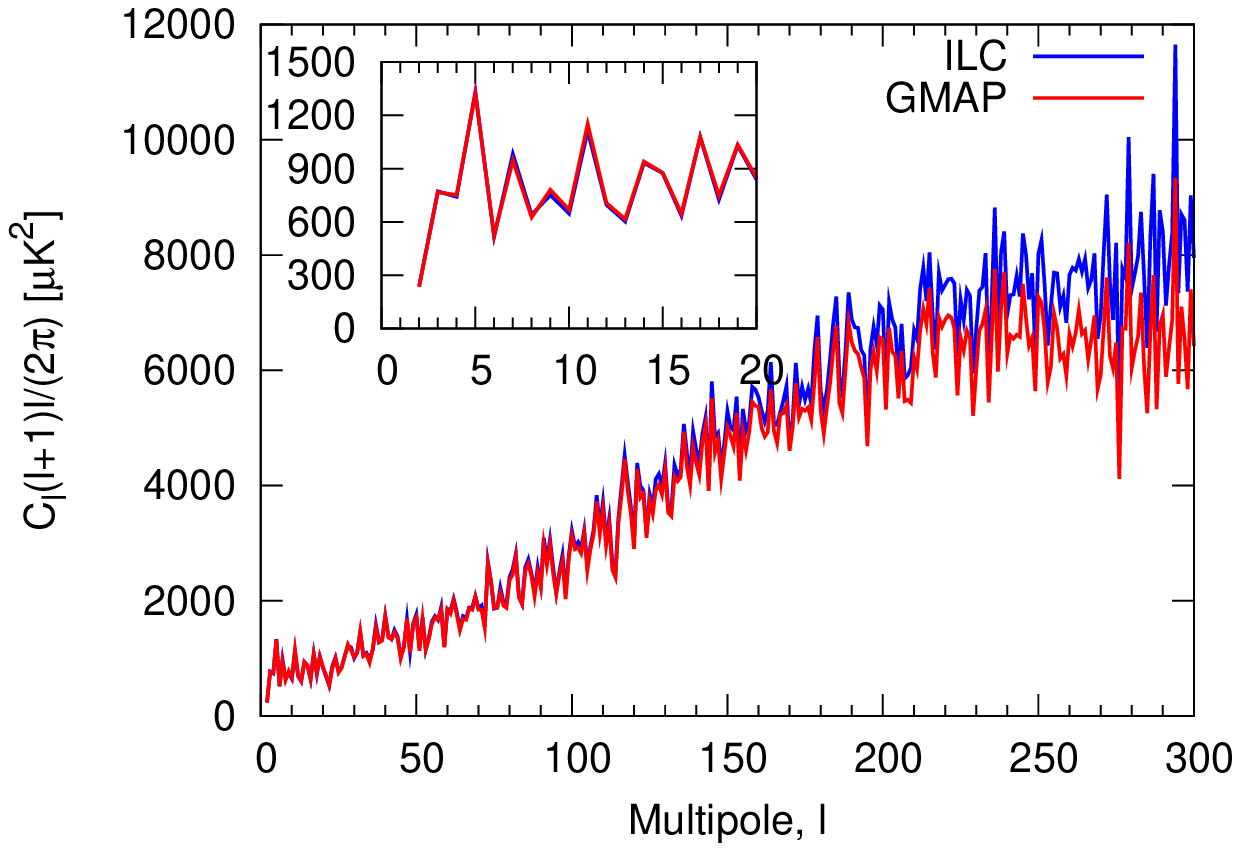}}
\caption{Full sky estimate of CMB Power spectrum  from  GMAP outside the G20 mask 
is shown in red line. The full sky estimate of WMAP's ILC power spectrum computed  from the same region 
is shown in blue line. Both spectra have been corrected for $1^\circ$ beam effect and pixel window effect.
In the inset we zoom in low multipole region. Both spectra match excellently with one another until $l\sim100$.\label{cl}}
\end{inlinefigure}

\section{Discussions \& Conclusion}
\label{Conclusion}
In this paper we have developed, validated and applied on $7$ year WMAP data a  
 global foreground minimization method  from CMB sky based upon the measure of non-Gaussian 
nature of the diffuse galactic foregrounds. The  GMAP obtained by this method matches 
well with the WMAP's  ILC map outside the G20 mask. The power spectrum from GMAP also
matches excellently with the power spectrum of ILC map until $l\sim 100$. At even higher $l$
we find less power than WMAP's ILC map. The  kurtosis value of 
the foreground minimized  map outside the G20 mask, $0.093 \pm 0.06$, the error is estimated 
from Monte-Carlo simulations of foreground removal method. This is only a $1.5\sigma$ effect  
implying  that the CMB sky  outside the G20 region is sufficiently clean so that the signals 
from this region may be interpreted to have cosmological origin consistent with  standard 
cosmological scenario. 

A crucial advantage of the method over the usual variance based ILC method is that the former does not possesses
any negative bias at low multipoles. The quadrupole moment estimated by us matches excellently
with the WMAP's ILC estimate. Our result shows that the problem of low  quadrupole moment persist in WMAP data. 
The quadrupole and octopole maps also show similar nature as the ones estimated from the WMAP's ILC map. The 
problem of quadrupole and octopole alignment using GMAP would be investigated in a future paper - however, given the 
similarity between GMAP and ILC map (at the least, outside the G20 mask) the alignment  likely to remain.  Another 
direction for future research would be to implement the method in the harmonic domain of the maps. The method can be 
generalized to estimate CMB angular power spectrum down to low angular scale following approaches described
in~\cite{Saha06} and ~\cite{Saha08}. Since  polarized foreground models are poorly known, an excellent 
research direction would be to apply our  
method on polarization sensitive CMB data released by WMAP  and PLANCK satellite mission~(\cite{Tauber01}).

Our method shows that it is possible to remove foregrounds from the CMB sky purely based upon non-Gaussian nature 
of non-cosmological signals. However, if there is any cosmological signal of non-Gaussian origin the method would 
interpret them as {\it foregrounds} and hence would try to eliminate the signal from the cleaned map. Therefore
comparing the GMAP with other foreground minimized maps  which do not explicitly rely upon the assumption of 
non-Gaussianity one may be able to isolate any possible primordial and/or secondary non-Gaussian signal. 
Estimating a Gaussian CMB map  using data from current and future generation
sensitive CMB experiments like PLANCK and CMBPol would play an important role to unfold this scientific information.  

We thankfully acknowledge useful discussions with Tarun Souradeep and  Jayanta Kumar Bhattyacharjee. 
Some of the results of this paper are obtained by using HEALPix 
software package~\citep{Gorski05}~{\footnote{The HEALPix distribution is publicly available from the website
http://healpix.jpl.nasa.gov.}}. We acknowledge use of LAMBDA data repository for our analysis.


\end{document}